# REQUIREMENTS VARIABILITY SPECIFICATION FOR DATA INTENSIVE SOFTWARE

Eman Muslah and Said Ghoul

Faculty of Information Technology, Research Laboratory on Bio-inspired Software Engineering,Philadelphia University, Amman, Jordan

## ABSTRACT

*Nowadays, the use of feature modeling technique, in software requirements specification, increased the variation support in Data Intensive Software Product Lines (DISPL) requirements modeling. It is considered the easiest and the most efficient way to express commonalities and variability among different products requirements. Several recent works, in DISPLs requirements, handled data variability by different models which are far from real world concepts. This,leaded to difficulties in analyzing, designing, implementing, and maintaining this variability. However, this work proposes a software requirements specification methodology based on concepts more close to the nature and which are inspired from genetics. This bio-inspiration has carried out important results in DISPLs requirements variability specification with feature modeling, which were not approached by the conventional approaches.The feature model was enriched with features and relations, facilitating the requirements variation management, not yet considered in the current relevant works.The use of genetics-based methodology seems to be promising in data intensive software requirements variability specification.*

## FUNDING

Philadelphia University, Research Project: Bio-inspired systems variability modeling project 12/2013

## 1.INTRODUCTION

Software Product Lines (which might be data intensive) requirements modeling is an approach to create more products that belong to a specific family for a specific domain from existing assets, which have the common characteristics either functional (services) or non-functional (data models) [1-6]. The most important aspect to take into consideration, when specifying requirements [23] of DISPL is determining the appropriate mechanism to model the variability in an efficient way [7, 28]. One of these mechanisms is feature model [8, 9, 10]. Its tree representation consists of a set of named features and relationships between them.

DISPLs are Software Product Lines requiring data as a major factor in their work, so they can handle large volume of data. This kind of SPL requires a data model to represent variability after fixing features [27, 29, 30].Software versioning [11, 12]is a management of software

                                                                    

modifications in a way that keeps general commonalities (functions) of software, satisfying new enhancements based on desired requirements, and facilitating the traceability process among different configurations of same software. Data versions [13, 14, 15, 16] are one of the data characteristics, which means from time to time, the need for a data change to cover new requirements.

A bio-inspired approach [17, 18]is a combination between the biological and the artificial life, in a way to enhance the artificial life through inspired from biological life characteristics. Because they success in solving many artificial problems, there is an increasing demand for these approaches such as neural networks and genetic algorithms, and the improvements it made in hardware sections [19, 20, 21].

The data model is the kernel part of a DISPL. Some authors [4, 6] handled variability in data models(relational, Entity relationship, ..) and customize products which are Data Base Management Systems (DBMS) for each model according to specific requirements. However, others authors [3, 5] handled, in different ways, the variability in Data itself.

But, despite the important scale of the research in the DISPL data variability modeling, the relevant current approaches are still far from real world concepts. Consequently, this hasgiven rise to weaknesses in data variability modelling that can be summarized through several aspects: These approaches have never handled variation in business domain (conceptual model) and in specific application (physical level). They just dealt with some variability in application families (logical model). The versions and revisions of data are not supported by current requirements specification feature modeling techniques, which leads to difficulties in data evolution and maintenance.

The previous limitations motivated this work for developing a methodology, close to real world, that supports data variability in DISPLs requirements specification. Its bio-inspired approach, based on genetics [32] engendered the data variability scope generalization on three meta levels: business domain, applications family, and specific application. This generalization has enriched the feature modelingformalism with new features (like version, revision, integrity constraints, …) and relations (like import, relation,selected explicitly, selected by axioms, ...). The use of genetics-based approach seems to be promising in software assets variability modeling including programs, data, etc. The data feature model is more natural and richer than the current ones.

## 2. RELATED WORKS

There are different approaches used to model variability, such as Feature Models, Orthogonal Variability Models and Decisions Models, but the most common and widely used one in SPLs, is Feature Models [24]. In the following, some relevant related works will be presented in order to identify significant open problems.

There is a diversity of researches that handled data variability based on feature models. In [25], the authors proposed a methodology for addressing data variability and data integration in SPL at the domain level, based on Unified Modeling Language (UML) standards. Their approach is divided into domain and application engineering levels. The initial step in the domain level is to analyze the presented requirements (to know which of them can be placed in the feature model as mandatory feature or optional feature, etc). The feature model is built to cover already presented requirements with respecting their constraints and dependencies.The application level is based on selecting the desired features form the general feature model to customize products based on customer needs. After that the data model instance is generated for presenting the data of specific configuration in a coherent way.

In [26], the authors proposed a variability modeling method for modeling variability of persistent data based on establishing variable data model that presents the variability in database objects (entities and attributes). This is achieved through the mapping process between persistency features and their data concepts. As result the real data model will created based on the desired features that customer had select.

Data aspect in software product lines was the focus in[3]. The authors proposed a technique based on two modules: one for presenting the features of product (core module) and the second one for presenting the features that customers want to have in their products (delta module). Different data configurations are generated through applying delta module on the core module as a modification process.

Khedri and Khosravi[5] have handled data variability in Multi Product Lines (MPL). They proposed a method for creating a universal feature model for MPL based on extracting features details from different SPL data models leading to variability.

Despite the important scale of the research in data variability modelling [3, 4, 5, 6, 27, 29, 30], the concepts of data version and revision have not been captured infeature models, which lead to complexity and problems in data maintenance. According to the current researches, the weakness of modeling data variability is obvious through several aspects.They have never modelled domain data requirement variability using feature model at the conceptual layer as well as logical layer. There is no flexible mechanism for handling data models modification that may occur during data life.The large gap between used concepts and real world leads to complexity of proposed methods and data variability maintenance. Consequently, the previous limitations motivated this research work.

# 3. A GENETICS-BASED REQUIREMENTS VARIABILITY SPECIFICATION IN DISPLS

## 3.1 Supporting Example

A data example that is used frequently in universities to describe Student and Course data will support the proposed requirements specification methodology. Using feature model, the Figure 1 shows three variations of the data Conceptual Models: Relational, Object-Oriented, and Entity Relationships. Only the Relational Conceptual Model is expended here. It presents the domain of the Student and Course data, consisting of St-Name, St-Avg, Course-Name, etc.

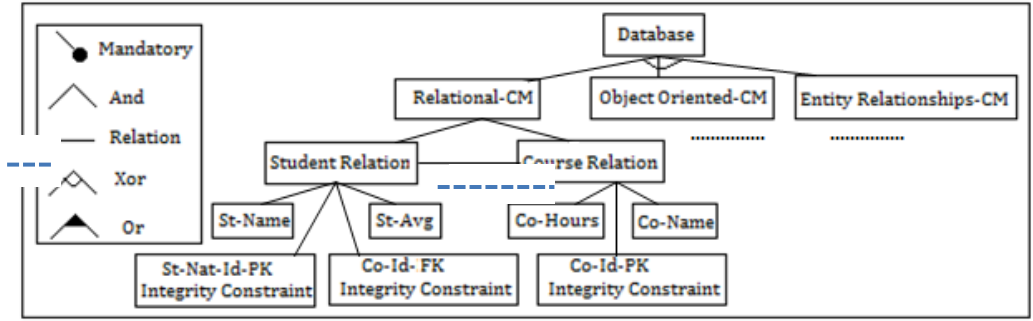

Figure 1: A Feature Model of Student and Course data.

## 3.2 A Methodology For Requirements Variability Specification In Displs

The proposed methodology uses feature modeling notations for requirements variability specification. It is shown in the Figure 2. It presents data feature met model with three sub meta models: Application Domain (Conceptual level, corresponding to genetics genome concept), Application Family (logical level, corresponding to genetics genotype concept), and Specific Application (physical level, corresponding to genetics phenotype concept).The starting is from the application domain variable requirements meta-model, then each application family variable and coherent meta-model will be derived, from the application domain meta-model, according to the application family variable requirements. At the end, specific application data model will be instantiated, from its family, according to its individual requirements. As a result any data-intensive application will be like a natural phenomenon.

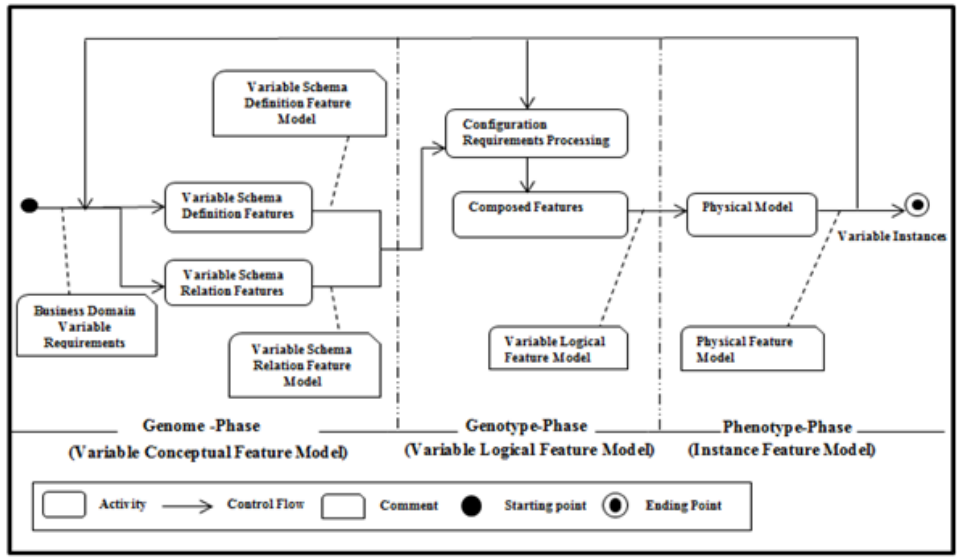

Figure 2: A Bio-inspired methodology for data requirements variability specification with feature model formalism.

The methodology is defined using UML notations Below, the methodology composing processes will be defined using EBNF notation [22] for presenting their structural aspects.

**a. Variable Data Feature meta Model (VDFM)**

The VDFM is composed of three sub models: Variable Conceptual Feature metaModel (VCFM), Variable Logical Feature meta Model (VLFM) and Variable Physical Feature Model (VPFM):

$$<VDFM> = \text{``}VDFM\text{''}: <Variable\ Data\ Feature\ Model\ name>\text{``};\text{''}$$
$$<VCFM>,<VLFM>,<VPFM>$$
$$\text{``}end\text{''}\ \text{``}VDFM\text{''} < Variable\ Data\ Feature\ Model\ name>\text{``};\text{''}$$

**b. Variable Conceptual Feature meta Model (VCFM)**

The VCFM deals with business domain (*like students management in any university*) variable requirements. It includes variable features that may be used to generate different logical met models according to the requirements of the application families. This meta model is composed of two sub meta models: Variable Schema Definition Feature meta Model (VSDFM) and Variable Schema Relation Feature meta Model (VSRFM):

$$<VCFM> = \text{``}VCFM\text{''}: <Variable\ Conceptual\ Feature\ Model\ name>\text{``};\text{''}$$
$$<VSDFM>,\ <VSRFM>$$

$$\text{``}end\text{''}\ \text{``}VCFM\text{''} < Variable\ Conceptual\ Feature\ Model\ name>\text{``};\text{''}$$

*b1. The Variable Schema Definition Feature meta Model (VSDFM)*is created according to business domain variable requirements for describing the definition of schema as features (Figure 3):

$$<VSDFM> = \text{``}VSDFM\text{''}: < variable\ schema\ definitions\ feature\ model\ name>\text{``};\text{''}$$
$$(<Relation\ features>,\ <relations>)^+;$$
$$\text{``}end\text{''}\ \text{``}VSDFM\text{''} < schema\ definitions\ feature\ model\ name>\text{``};\text{''}$$

Relation features are used to present the relations between data.They are composed of version features and their relations to denote the variation between them:

$$< Relation\ features> = \text{``}Relation\text{''}: <Relation\ features\ name>;$$
$$(<Version\ features>,\ <relations>)^+;$$

Version features (of a relation) are used to present the versions (alternatives variations) that may be created during data life and can have different names indicating their semantics. They are composed of revision features and their relations to denote the variation between them:

$$< Version\ features> = \text{``}Version\text{''}: <Version\ features\ name>;$$
$$(<Revision\ features>,\ <relations>)^+;$$

Revision features (of a version) are used to present the revisions for handling the modifications that may occur in one version through the evolution operations.They are composed of Field Definition and Integrity Constraint (IC) Definition features and their relations:

> *< Revision features > = "Revision" : <Revision features name>;*
> *((<Field Definition features>,*
> *[< IC Definition features>]) ⁺,*
> *<relations>)⁺;*

• *Field Definition features* are used to present the fields of schema of a data and through what evolution operation they are produced (Add field, Delete field, or Modify field)and their relations to denote the variation between them:

> *< Field Definition features>="Field Definition" : < Field Definition features name>;*
> *(("Add field" (<attributes>, <relations>) ⁺,*
> *["Delete field" (<attributes>, <relations>) ⁺],*
> *["Modify field" (<attributes>, <relations>) ⁺;*

*Add field feature, Delete field feature,* and *Modify field feature*are used to denote add, delete, and modify specified fields operations on data schema. *Attributes* formal definition will be presented once at the end of this section.

• *IC Definition features*are used to present the integrity constraints of schema and through what evolution operation they are produced (Add IC, Delete IC, or Modify IC)and their relations to denote the variation between them:

> *< IC Definition features >="IC Definition" : < IC Definition features name>;*
> *(("Add IC" (<attributes>, <relations>) ⁺,*
> *["Delete IC" (<attributes>, <relations>) ⁺],*
> *["Modify IC" (<attributes>, <relations>) ⁺)⁺,*
> *<relations>) ⁺;*

*Add IC features, Delete IC feature, and Modify IC feature*are used to denote add, delete, and modify specified integrity constraints operations on data schema. Attributes are used to denote the name of fields and integrity constraints:*<attributes>* = *(<Attr_name>:<Attr_value>)⁺*. The relations formal definition (used in the above definitions) is specified by: *and | xor | or| mandatory| empty*

*b2. The Variable Schema Relation Feature meta Model (VSRFM)*is created according to domain variable requirements beside VSDFM. This feature met model deals with the relations that may occur between revisions in different versions (in one or more data relations). It's separated from VSDFM for understand ability purposes (Figure 4):

> *<VSRFM>= "VSRFM": < Variable Schema Relation feature model name >";"*
> *(<Relation features>, <Relations>) ⁺*
> *"end" "VSRFM" <Schema Relation feature model name >";"*

*Relation features* and *Relations* are defined previously. *Revision features* (composing the relation features) handle the relations that may occur between revisions in one or more data relations (Imply, Exclude, and Import):

< Revision features >=“Revision” :   < Revision features name >;
    (([“Imply” (<attributes>, <relations>)$^+$],
                        [“Exclude” (<attributes>, <relations>)$^+$],
[“Import” (<attributes>, <relations>)$^+$])$^+$, <Relations >)$^+$;

A feature F1 *implies* a feature F2 if a revision holding F1 must hold F2 too. A feature F1 *excludes* a feature F2 if a revision holding F1 must not hold F2.A feature F1 *imports* a feature F2 if a revision1 needs feature from revision2.

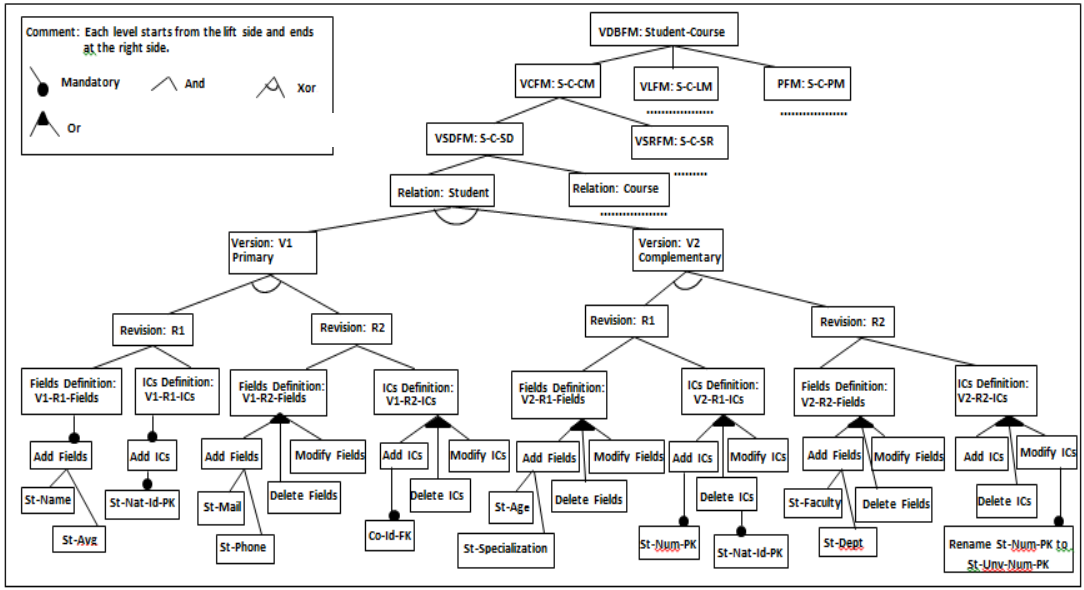

Figure 3: Variable Schema Definition Feature meta Model of Student data.

### c. Variable Logical Feature meta Model (VLFM)

Each applications family (*like students registration in any university*) requires a coherent and meaningful subset of features selected from the previous VCFM (*i.e. Students management in any university*). This subset is called a conceptual model configuration or simply a logical model. It is composed by selecting needed features and/or rejecting unwanted ones:

  <VLFM>= “VLFM”: < Variable Logical feature model name >“;”
    (<Application Family needs features>)$^+$; //In data versions conforming to feature relationships
    [(“VersionLM” : <VersionLM features name>, <relations>)$^*$]; //New versions may be created
    [(“Revision LM” : <Revision LM features name>, <relations>)$^*$]; //New revisions may be created
  “end” “VLFM” <Variable Logical feature model name >“;”

Version features are used to define the versions of logical models that might be created during database life and according to the applications family needs. Revision features are used to represent the revisions of logical models for handling the modifications that may occur in one logical model version. Application family needed features are used to fix applications family requirements according to the VCFM. These features are composed in a coherent way:< *Application family needs features* > = <*VCFM name*>:(<*feature*>,<*feature*>)+.The application family needs lead to select features through select explicitly relation. The implicitly selected features (selected by VSRFM relations), required by an applications family, are gathered from the VSDFM features according to VSRFM features in a coherent way.*Relations* are used to deal with variation of features:<*relations*> = *and* | *xor* | *or| mandatory.*The figure 5 shows a VLFM generated by the following family selection program:

> *VLFMStV1-CsV2*
> *{ //Application family needs. Input VCFM. Output VLFM*
> *S-C-SD: Student>V1-primary, Course>V2      //selected features*
>   *// The obtained VLFM will be coherent with feature relationships: imply, exclude, etc.*
> *}*

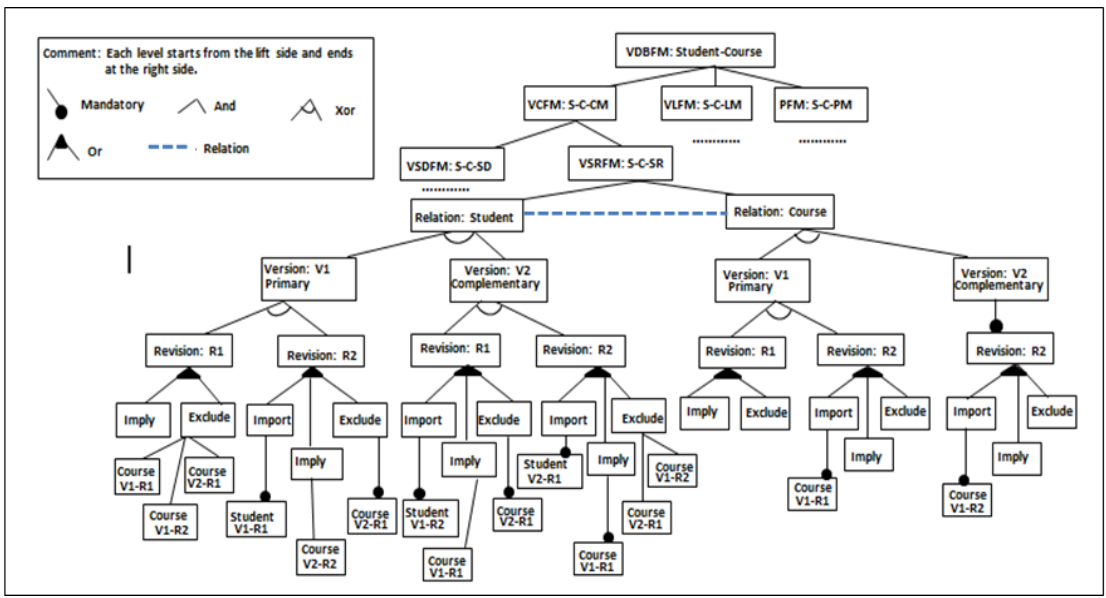

Figure 4: Variable Schema Relation Feature metaModel of Student and Course data.

## d. Instance ___ ture Model (IFM)

Each specific application (*like students registration at Philadelphia University*) requires a coherent and meaningful subset of features selected from the previous VLFM (*i.e. Students registration in any university*). This subset is called an application instance model or a physical model. It is composed by selecting needed features and/or rejecting unwanted ones:

> <*IVFM*>= "IVFM":<*Instance variable feature model name* >";"
>   .(<*Application  needs features*>)+; //In data revisions conforming to feature relationships
>   [("Revision IVM" : <*Revision IVM features name*>, <*relations*>)*]; //New revisions may be created
> "end" "IVFM" <*Variable Logical feature model name* >";"

Revision features are used to represent the revisions of instance models for handling the modifications that may occur in one instance model. Application needs features are used to fixer specific application needs according to the VLFM. These features are composed in a coherent way: *Application needs features* $> = (<feature>,<feature>)^+$. The figure 6 shows a VIFM generated by the following application instance selection program:

> *VIFM StV1R2-CsV2R2*
> *{ //An Application needs. Input VLFM. Output VIFM*
> *StV1-CsV2:  Student>V1-primary>R2; // the feature Course>V2>R2 will be selected*
> *//automatically (implied by Student>V1-*
> *primary>R2.*
> *}*

**e. Instance Data schema (IDS)**

Each VIFM (*like students registration at Philadelphia University*) might be used to generate specific application data schema (*i.e. Students registration at Philadelphia University data schema*). This schema defines the real data requirements specification model for that application.

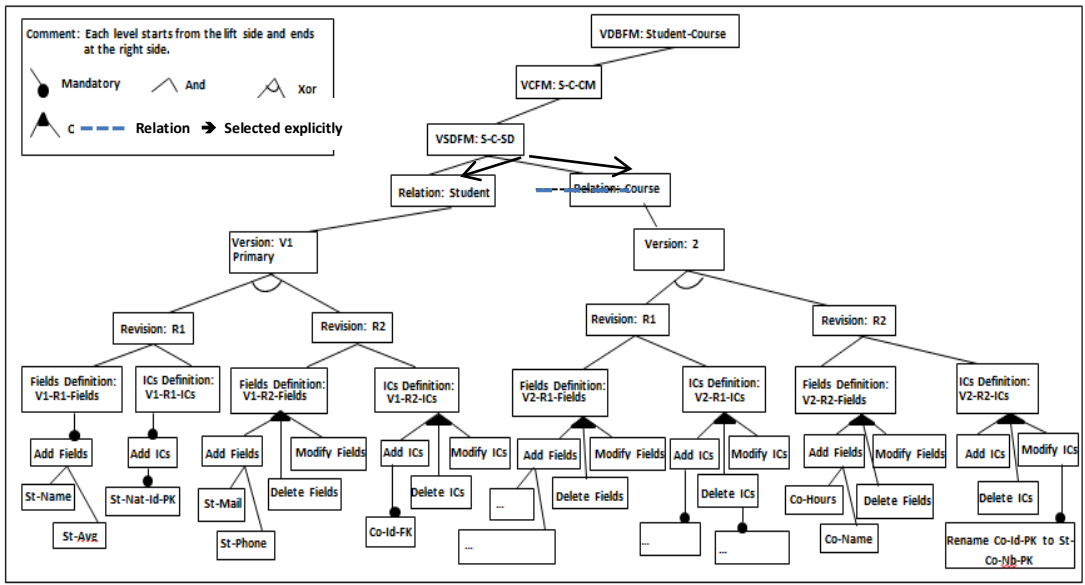

Figure 5: *VLFM StV1-CsV2*

The figure 7 shows a IDS generated by the following application instance selection program:

> *IDS StV1R2-CsV2R2*
> *{ // Input VIFM. Output IDBS*
> *StV1R2-CsV2R2:  Student>V1-primary>R2, Course>V2>R2. //selected features.*
> *  // When an IDS evolves with revisions, new schemas may be generated accordingly.*
> *}*

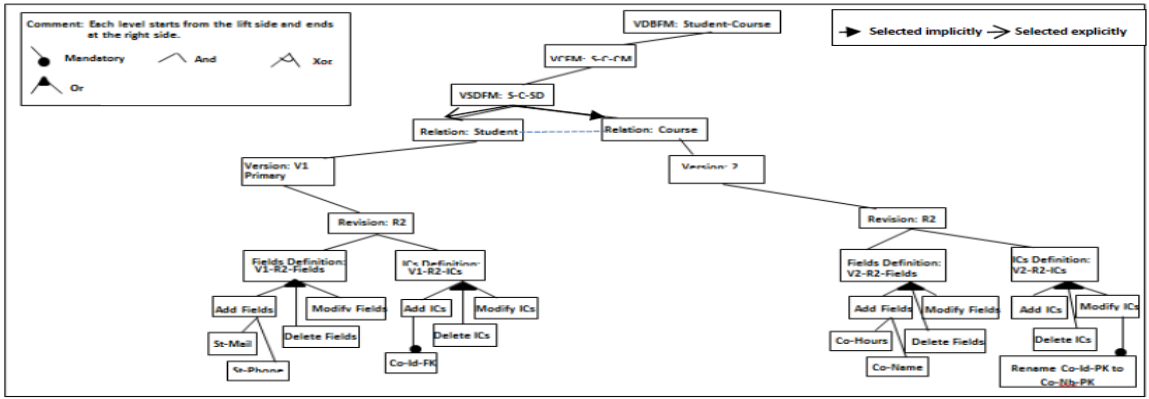

Figure 6: *VIFM  StV1R2-CsV2R2*

> *Student (St-Name string, St-Avg float, St-mail string , St-Pone number, St-Nat-Id-Pk constrain, Co-Id-Fk constraint)*
> *Course (Co-Name string, Co-Hour string, Co-Nb-Pk constraint)*

Figure 7: The IDS  StV1R2-CsV2R2.

# 4. CONCLUSION

This approach is recommended to be used in any variable data requirements specification area for data intensive software, like software product lines, multiple software product lines, and data intensive product lines.A comparison between the proposed requirements specification methodology (for data variable requirements),with its new concepts, and the others similar works with their traditional concepts was achieved according the following common criteria: (1) real world concepts based, (2) Broad Variability meta modeling, (3) version and revision based, (4) Feature model enhancement, (5) Application data schema based on specific needs, and (6) data maintenance decreased efforts.All the above studied similar works support the criteria 5 and 6, but, they are not dealing with the first four ones. However, the proposed approach covers all of them.

Based on the study, previously presented in this work,of data variability in DISPLs requirements specification, several insufficiencies were identified in current approaches. They are mainly due the large gap separating them from the real world. This leaded to weaknesses in their methodologies as well as in their supporting methods:  poor variability meta modeling specification and relatively poor supporting methods.

In this work, a close to real world methodology(based on genetics) has leaded to enhancing the above insufficiencies:(1) Data requirements variability specification with meaningful and complete meta modeling levels and (2) supporting methods enrichment like feature modeling extension, variability enrichment with version and revision techniques, and automatic specific application data requirements generation. These enhancements facilitate the data requirements evolution and allow tracking and reversing the requirements evolution of data. The variability of the data requirements definition was completely presented through variability of fields and integrity constraints in a uniform way. When an application family selects the desired requirements features, it will keep them in a coherent way, due to the relations that reflect consistency between versions and revisions. The automatic business domain requirements translation into a VCFM, the integration of the data modeling variability (like relational, object-oriented, Object-relational, …), the variability of specific application data requirements, the continuous data variability requirements engineering, the variable data requirements reverse engineering, and the evolution of applications programs accompanying the evolution of their data requirements are real challenges.

## REFERNCES

.